\documentclass[letterpaper, 10 pt, conference]{ieeeconf}  

\IEEEoverridecommandlockouts                              

\usepackage{color}

\usepackage{times}
\usepackage{epsfig}
\usepackage{graphicx}
\usepackage{tabularx}
\usepackage{amsmath}
\usepackage{amssymb}
\usepackage{bbm}
\usepackage[misc]{ifsym}
\usepackage[export]{adjustbox}
\let\labelindent\relax
\usepackage{enumitem}

\usepackage[ruled,vlined,linesnumbered]{algorithm2e}
\usepackage{graphicx} 

\usepackage{subfigure}
\usepackage{tikz}
\usepackage{verbatim}
\usepackage{booktabs}
\usepackage{multirow}
\usepackage{makecell}
\usepackage{authblk}
\usepackage{hyphenat} 

\usepackage{pgf}
\usepackage{algpseudocode}
\usetikzlibrary{arrows,automata}

\usepackage{graphbox}

\makeatletter
\newcommand{\removelatexerror}{\let\@latex@error\@gobble}
\makeatother

\makeatletter 
  \newcommand\figcaption{\def\@captype{figure}\caption} 
  \newcommand\tabcaption{\def\@captype{table}\caption} 
\makeatother

\makeatletter
\let\NAT@parse\undefined
\makeatother
\usepackage[colorlinks]{hyperref}  

\title{
How Can Robots Trust Each Other For Better Cooperation? \\
A Relative Needs Entropy Based Robot-Robot Trust Assessment Model}

\author{Qin Yang \and Ramviyas Parasuraman \thanks{$^{*}$ The authors are with the Heterogeneous Robotics Research Lab (HeRoLab), Department of Computer Science, University of Georgia, Athens, GA 30602, USA. 
Email: {\it \{qy03103,ramviyas\}@uga.edu}. }}

\begin{document}

\newtheorem{definition}{Definition}
\newtheorem{theorem}{Theorem}
\newtheorem{lemma}{Lemma}
\newtheorem{proposition}{Proposition}
\newtheorem{property}{Property}
\newtheorem{observation}{Observation}
\newtheorem{corollary}{Corollary}

\maketitle
\thispagestyle{empty}
\pagestyle{empty}

\begin{abstract}
Cooperation in multi-agent and multi-robot systems can help agents build various formations, shapes, and patterns presenting corresponding functions and purposes adapting to different situations. Relationship between agents such as their spatial proximity and functional similarities could play a crucial role in cooperation between agents. Trust level between agents is an essential factor in evaluating their relationships' reliability and stability, much as people do. This paper proposes a new model called Relative Needs Entropy (RNE) to assess trust between robotic agents. RNE measures the distance of needs distribution between individual agents or groups of agents. To exemplify its utility, we implement and demonstrate our trust model through experiments simulating a heterogeneous multi-robot grouping task in a persistent urban search and rescue mission consisting of tasks at two levels of difficulty. The results suggest that RNE trust-Based grouping of robots can achieve better performance and adaptability for diverse task execution compared to the state-of-the-art energy-based or distance-based grouping models.
\end{abstract}


\section{Introduction}

Trust describes the interdependent relationship between agents \cite{swinth1967establishment}, which can help us better understand the dynamics of cooperation and competition, the resolution of conflicts, and the facilitation of economic exchange \cite{lewicki2006models}. In different computing fields, trust is different based on various perspectives. However, in computational agents, these models do not involve the trustor's characteristics but focus on forming trust based on past behavior \cite{cho2015survey}.

In \textit{Automation}, authors in \cite{lee2004trust} describe trust in human-automation relationships as the attitude that an agent will help achieve an individual’s goals in a situation characterized by uncertainty and vulnerability. The trustor here is a human and the trustee usually is the automation system or an intelligent agent like robot. Those systems' primary purpose is to assess and calibrate the trustors' trust beliefs reflecting the automation system's ability to achieve a specific task.

In \textit{Computing and Networking}, the concept of trust involves many research fields, such as artificial intelligence (AI), human-machine interaction, and communication networks \cite{cho2010survey,sherchan2013survey,wang2020survey}. They regard an agent's trust as a subjective belief which represents the reliability of the other agents' behaviors in a particular situation with potential risks. Their decisions are based on learning from the experience to maximize its interest (or utility) and minimize risk \cite{cho2015survey}. Especially in the social IoT domain \cite{chen2015trust,mohammadi2019trust}, trust is measured based on the information accuracy and agents' intentions (friendly or malicious) according to the current environment to avoid attacks on the system.

\begin{figure}[tbp]
\centering
\includegraphics[width=0.49\textwidth]{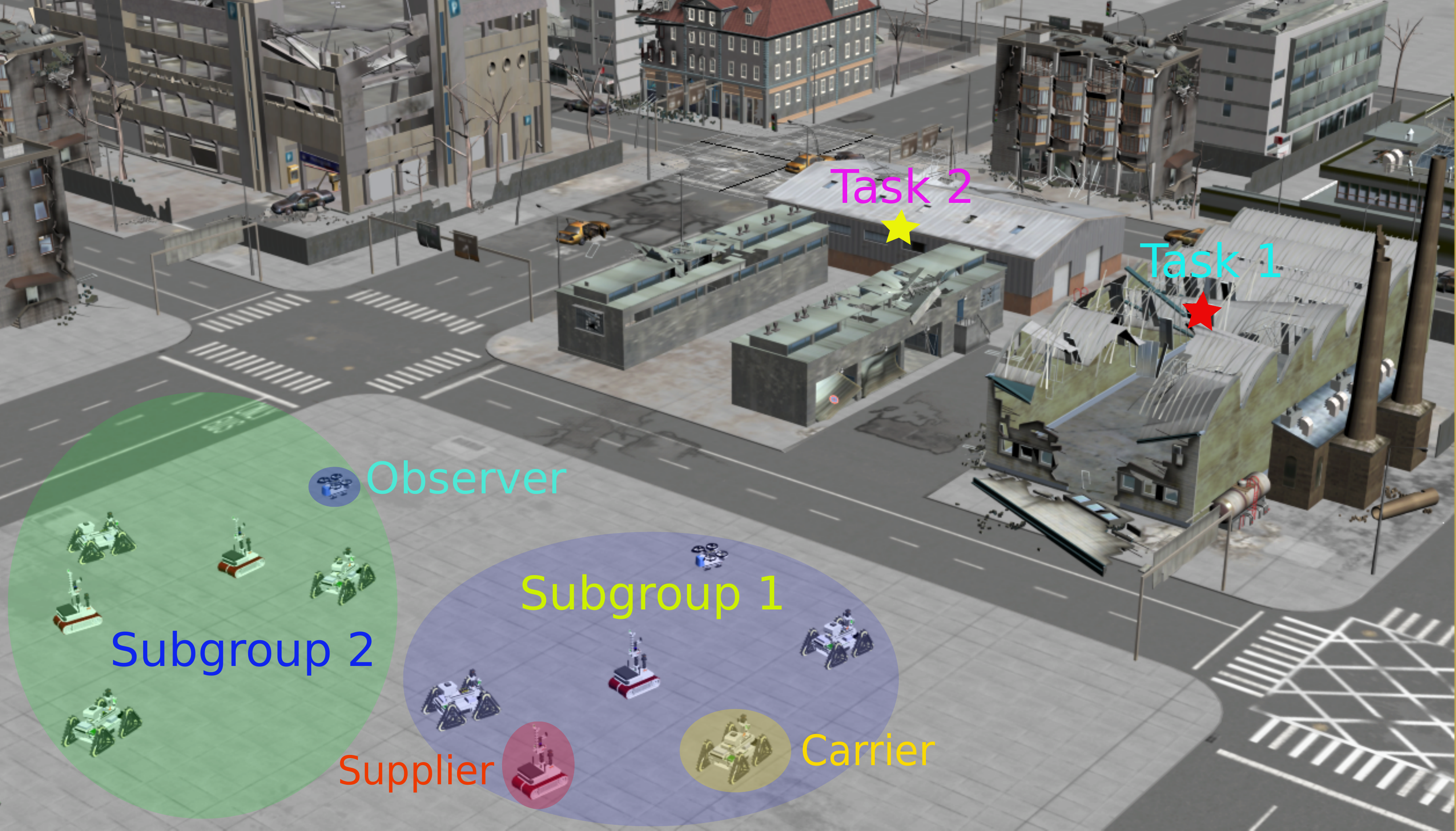}
\caption{Illustration of two heterogeneous robot teams (ground robots + aerial robots) cooperatively achieving tasks of two different difficulty levels in a post-nuclear leak rescue mission.}
\label{fig: overview}
\vspace{-4mm}
\end{figure}

In \textit{Multi-Robot Systems (MRS)}, trust by one agent on another agent reflects the trusting agents' mission satisfaction, organizational group member behaviors and consensus, and task performance at the individual level. Moreover, trust has been associated with system-level analysis, such as reward and function at the organized group structure in a specific situation, the conflict solving of heterogeneous and diverse needs, and the system evolution through learning from interaction and adaptation. 
On the other hand, the act of trusting by an agent can be conceptualized as consisting of two main steps: 1) \textit{trust evaluation}, in which the agent assesses the trustworthiness of potential interaction partners; 2) \textit{trust-aware decision-making}, in which the agent selects interaction partners based on their trust values \cite{yu2013survey}. Also, assessing the performance of agent trust models is of concern to agent trust research \cite{fullam2005specification}.

Considering the interactions between human agents and artificial intelligence agents like human-robot interaction (HRI), building stable and reliable relationships is of utmost importance in MRS cooperation, especially in adversarial environments and rescue missions \cite{yang2020needs}. Fig. \ref{fig: overview} illustrates two heterogeneous robot teams cooperating to achieve two different challenging tasks in a post-nuclear leak rescue mission. In such multi-agent missions, appropriate trust in robotic collaborators is one of the leading factors influencing HRI performance \cite{khavas2020modeling}. In HRI, factors affecting trust are classified into three categories \cite{hancock2011meta}: 1) \textit{Human-related factors} (i.e., including ability-based factors, characteristics); 2) \textit{Robot-related factors} (i.e., including performance-based factors and attribute-based factors); 3) \textit{Environmental factors} (i.e., including team collaboration, tasking, etc.).

Although modeling trust has been studied from various perspectives and involves many different factors, it is still challenging to develop a conclusive trust model that incorporates all of these factors. 
Therefore, future research into trust modeling needs to be more focused on developing general trust models based on measures other than the countless factors affecting trust \cite{khavas2020modeling}. 
Specifically, no work from the literature considered trust in socio-intelligent systems  from an artificial intelligent agent-perspective (robots) to model trust with agents having similar needs and interests to the best of our knowledge.
Balch \cite{balch1997social} laid the foundations in introducing behavioral difference and social entropy metrics to evaluate diversity in societies of mechanically similar agents from the agent's behaviors perspective in an MRS. However, it only looks from a diversity point of view and does not establish a substantial and clear connection between the agent's behaviors and motivations, especially trust.

To bridge this important gap, in this paper, we introduce a general agent-agent trust model based on \textit{Relative Needs Entropy (RNE)}. Here, we extend our previous work \cite{yang2019self}, where we defined needs hierarchy for robots  similar to Maslow's law of human needs. Using the current state of the needs of the agents, it is possible to devise a relative needs entropy value, which can help measure social trust between agents. We verify this trust model in a simulated urban search and rescue (USAR) mission to achieve multi-robot grouping based on trust level within the robots in a group. We analyze the relative mission performance against  state-of-the-art methods which use energy capability (battery level) \cite{yang2020hierarchical} or the spatial proximity (i.e., the inter-distance) between the robots for distributed task allocation \cite{parasuraman2018multipoint}. 
The specific contributions made in this paper are as follows:
\begin{enumerate}
    \item \textbf{RNE-based robot-robot\footnote{We use the terms robot and agent interchangeably} and multi-robot group trust models}
Considering intelligent agents representing different needs in interaction, especially for the heterogeneous MRS, we introduce a new RNE-based trust assessment model to describe trust levels between agents or groups. It presents the similarity of their needs.

\item \textbf{RNE trust based heterogeneous multi-robot grouping mechanism}
We propose a hierarchical approach for RNE trust-based formation of groups in a heterogeneous MRS. This method will first form homogeneous agent subgroups, then hierarchically merges according to a specific sequence until forming the final heterogeneous groups' combination.

\end{enumerate}

We provide examples of the proposed trust model and demonstrate its effectiveness in heterogeneous multi-robot grouping for a simulated urban search and rescue (USAR) mission, where the robots should form two specific groups to finish two tasks at easy and hard difficulty levels based on the environment. We hypothesize that RNE trust-based group formation would achieve better cooperation (lower costs and higher group rewards) by balancing and aligning their motivations (i.e., agent needs). 

\section{Background and Preliminaries}

This section briefly reviews the \textit{relative Entropy}, and our previous study on \textit{robot needs hierarchy}. The relative definitions and notations are described below.

\subsection{Relative Entropy}
The \textit{relative entropy} or the Kullback Leibler distance is a measure of distance between two distributions. In statistics, it arises as an expected logarithm of the likelihood ratio. The relative entropy $\mathop{D(p||q)}$ is a measure of the inefficiency of assuming that the distribution is $q$ when the true distribution is $p$ \cite{cover1991entropy}. The corresponding mathematical definition is shown as Eq.~\eqref{relative_entropy}.
\begin{equation}
\begin{split}
    \mathop{D(p||q)} & = \sum_{x \in X} p(x) \log \frac{p(x)}{q(x)} \\
                     & = \mathop{E_p} \log \frac{p(X)}{q(Y)}.
    \label{relative_entropy}
\end{split}
\end{equation}

Furthermore, the \textit{mutual information} \cite{cover1991entropy} represents a measure of the amount of information that one random variable contains about another random variable and the reduction in the uncertainty of one random variable. Supposing we consider two random variables $X$ and $Y$ with a joint probability mass function $p(x,y)$ and marginal probability mass function $p(x)$ and $p(y)$. The \textit{mutual information} $\mathop{I(X;Y)}$ is the \textit{relative entropy} between the joint distribution and the product distribution $p(x)p(y)$ as Eq.~\eqref{mutual_information}.
\begin{equation}
\begin{split}
    \mathop{I(X;Y)} & = \sum_{x \in X} \sum_{y \in Y} p(x,y) \log \frac{p(x,y)}{p(x)p(y)} \\
                    & = \mathop{D(p(x,y)||p(x)p(y))} \\
                    & = \mathop{E_{p(x,y)}} \log \frac{p(X,Y)}{p(X)P(Y)}.
    \label{mutual_information}
\end{split}
\end{equation}

\subsection{Robot Needs Vector}

In \textit{Robot Needs Hierarchy} \cite{yang2019self,yang2020hierarchical}, the agent's needs  expressed as the \textit{Needs Expectation}, which can be calculated through its behaviors' weight and corresponding safety probability based on the data of perception and communication. The individual safety needs (Eq.~\eqref{safety_need}) are the precondition for calculating the basic needs (Eq.~\eqref{basic_need}). Similarly, only after fitting the safety and basic needs can consider its capability needs (Eq.~\eqref{capability_need}) and the teaming needs (Eq.~\eqref{teaming_need})
\begin{equation}
\begin{split}
    Safety~~Needs:~~ N_{s_{t_i}} = \sum_{i=1}^{s_{t_i}} W_{i} \cdot \mathbb{P}(W_{i}|P, C); \label{safety_need}
\end{split}
\end{equation}
\vspace{-2mm}
\begin{equation}
\begin{split}
    Basic~~Needs:~~ N_{b_i} = \sum_{i=1}^{b_i} W_{i} \cdot \mathbb{P}(W_{i}|P, C, N_{s_{t_i}}); \label{basic_need}
\end{split}
\end{equation}
\vspace{-2mm}
\begin{equation}
\begin{split}
    Capability~Needs: N_{c_i} = \sum_{i=1}^{c_i} A_i \cdot \mathbb{P}(A_i|T, P, N_{b_i}); \label{capability_need}
\end{split}
\end{equation}
\vspace{-2mm}
\begin{equation}
\begin{split}
    Teaming~~Needs:~~ N_{t_i} = \sum_{i=1}^{n} U_i \cdot \mathbb{P(}U_i | P, C, N_{c_i}); \label{teaming_need}
\end{split}
\end{equation}

\textbf{Here,} $P$ and $C$ represent the data of the agent's perception and communication separately; $T$ represents the task requirement space; $U_i$ represents the utility value of agent $i$ in the group; $W$ represents corresponding weights; $A$ represents the level of the agent's corresponding capabilities based on task requirements; $b_i$, $s_{t_i}$, and $c_i$ represent the size of agent $i$'s basic, safety, and capability needs solution space respectively; $n$ represents the number of agents in the group.

Through the above analysis, we adopt \textit{Utility Theory} \cite{fishburn1970utility,kochenderfer2015decision} to define the agent's fourth level needs -- \textit{Teaming Needs} (Eq. \eqref{teaming_need}), representing higher-level needs for an intelligent agent. It can be regarded as a kind of motivation or requirement for cooperation achieving specific goals or tasks to satisfy the individual or group's certain expected utilities. 

\subsection{Relative Needs Entropy}
\label{sec:rne}
Similar to the information entropy, we define the needs entropy as the difference or distance of needs distribution between agents in a specific scenario for an individual or groups. Here, needs of the robots are regarded as their motivations. From a statistical perspective, the RNE can be regarded as calculating the similarity of high-dimensional samples from the robot needs vector. \textbf{A lower RNE value means that the trust level between agents or groups is higher because their needs are well-aligned and there is low difference (distance) in their needs distributions}. Similarly, a higher RNE value will mean that the needs distributions are diverse, and there exists a low trust level between the agent or groups because of their misalignment in their motivations, which are similar to each other. More specifically, we formalize the RNE based trust model on three different situations in Sec.~\ref{3}.

\section{Proposed RNE Based Trust Models}
\label{3}

\newtheorem{proCoro}{\textbf{Definition}}
\begin{proCoro}[Trust between Agents]
\label{trust1}
Supposing the needs' vectors of R$_1$ and R$_2$ are N$_{R_1}$(n$_{11}$, \dots, n$_{1j}$) and N$_{R_2}$(n$_{21}$, ..., n$_{2j}$), where $j$ is the number of specific needs (categories) in the needs space. Then, through the corresponding weight vector W($w_1$, \dots, $w_j$), we get the needs' distribution of two agents are D$_{R_1}$(d$_{11}$, \dots, d$_{1j}$) and D$_{R_2}$(d$_{21}$, ..., d$_{2j}$) respectively. We can present the RNE based Trust value from R$_1$ to R$_2$ as Eq. \eqref{agents_trust}. Here, d$_{1k}$ and d$_{2k}$ are calculated as Eq. \eqref{agents_needs_distribution}. (j, k $\in$ $Z^+$)
\begin{equation}
    \mathop{\mathbb{T}(R_1||R_2)} = \sum_{k=1}^{j} D_{R_{1_k}} \cdot \log \frac{D_{R_{1_k}}}{D_{R_{2_k}}}
\label{agents_trust}
\end{equation}
\begin{equation}
    D_{R_{1/2}} \ni \mathop{d_{1k/2k}} = n_{1k/2k} \cdot w_k / \sum_{k=1}^{j} (n_{1k/2k} \cdot w_k)
\label{agents_needs_distribution}
\end{equation}
\end{proCoro}

Note, $ \mathop{\mathbb{T}(R_1||R_2)} \not\equiv \mathop{\mathbb{T}(R_2||R_1)} $ since we consider relative entropy. That is, robot $R_1$'s trust on the robot $R_2$ is not necessarily reciprocal (equal to the robot $R_2$'s trust on $R_1$) because the reference "true" distributions are different in both cases. Moreover, the robot's information on other robot's needs expectations of current levels need not be accurate because of information uncertainty. This makes the RNE trust model applicable in most realistic scenarios where robots rely on perception and communication to gather information about other robots in the scenario.

\paragraph*{Example 1}
\label{example1}
Let a robot has three needs value to represent its safety, energy, and health levels. Assume the needs' vectors of three robots $R_1$, $R_2$, and $R_3$ are [86, 120, 30], [20, 30, 10], and [80, 115, 25], respectively. The corresponding weight vector is [6, 4, 2]\footnote{Due to the difference in individual needs and situations, the weight vector might be different correspondingly. Here, we consider equal weights.}, which represents the priority each needs level belongs to. According to Eq. \eqref{agents_needs_distribution}, we can get their needs' distributions as D$_{R_1}$(0.4884, 0.4545, 0.0568), D$_{R_2}$(0.4615, 0.4615, 0.0769), and D$_{R_3}$(0.4848, 0.4646, 0.0505). Then,
\begin{equation}
\begin{split}
    \mathop{\mathbb{T}(R_1||R_2)} = & 0.4886 \cdot \log \frac{0.4886}{0.4615} + 0.4545 \cdot \log \frac{0.4545}{0.4615} + \\
    & 0.0568 \cdot \log \frac{0.0568}{0.0769} = 0.00373
\label{agents_trust_example1}
\end{split}
\end{equation}
\begin{equation}
\begin{split}
    \mathop{\mathbb{T}(R_1||R_3)} = & 0.4886 \cdot \log \frac{0.4886}{0.4848} + 0.4545 \cdot \log \frac{0.4545}{0.4646} + \\
    & 0.0568 \cdot \log \frac{0.0568}{0.0505} = 0.0005
\label{agents_trust_example2}
\end{split}
\end{equation}

By calculating the RNE values between $R_1$ and $R_2$ Eq. \eqref{agents_trust_example1}, and $R_3$ Eq. \eqref{agents_trust_example2}, the trust values $\mathop{\mathbb{T}(R_1||R_3)} < \mathop{\mathbb{T}(R_1||R_2)}$. Therefore, as per the definition of RNE in Sec.~\ref{sec:rne}, robot $R_1$ has a higher trust level with $R_3$ compared to its trust on $R_2$ because robot $R_1$ shares similarities with the robot $R_3$ in terms of needs distributions and hence the RNE value is lower.

\begin{proCoro}[Trust between Agent and Groups]
\label{trust2}
Supposing the needs' vector of the agent $R$ is $N_R$, and the needs matrix of the group $R_G$ is ($N_{1}$, \dots, $N_{i}$), where i represents the number of robots in the group with corresponding weights W($w_1$, \dots, $w_j$) for the needs categories. According to Def. \ref{trust1}, we can represent the needs' distribution vector of R as D$_R$. Considering $R_G$ needs' distribution vector as D$_{R_G}$ Eq. \eqref{agent_group_needs_distribution}, the RNE Trust value from the robot $R$ to the group R$_G$ can be represented as Eq. \eqref{agent_group_trust}. (i,j $\in$ $Z^+$)
\begin{equation}
\label{agent_group_needs_distribution}
\begin{split}
\mathop{D_{R_G}} & =
\left[
\begin{matrix}
N_1 \\ 
N_2 \\ 
\vdots \\ 
N_i \\ 
\end{matrix} 
\right]
\end{split}_{i \times j}
. \times
\left[
\begin{matrix}
w_1 \\ 
w_2 \\ 
\vdots \\ 
w_j \\ 
\end{matrix} 
\right]^T
./
\left[
\begin{matrix}
\sum_{k=1}^{i} (N_k\cdot W_k)
\end{matrix}
\right]_{j \times 1}
\end{equation}
\begin{equation}
    \mathop{\mathbb{T}(D_R||D_{R_G})} = \sum_{k=1}^{j} D_{R_k} \cdot \log \frac{D_{R_k}}{D_{R_{G_k}}}
\label{agent_group_trust}
\end{equation}
\end{proCoro}

\paragraph*{Example 2} 
\label{example2}
In addition to the Robot $R_1$ from Example 1, assume we have two robot groups $R_{G_1}$ and $R_{G_2}$ with two robots in each group. Their needs' matrices are $\{$[82, 114, 24]; [79, 117, 23]$\}$ and $\{$[40, 56, 12]; [56, 48, 15]$\}$ respectively. The weight vector of needs is the same as in Example 1. Through the Eq. \eqref{agent_group_needs_distribution}, we get their needs' distributions as D$_{R_{G_1}}$ (0.4869, 0.4657, 0.0474) and D$_{R_{G_2}}$ (0.5507, 0.3977, 0.0516). Then we can get the RNE values between robot $R_1$ and group $R_{G_1}$ Eq. \eqref{agents_group_trust_example1}, and $R_{G_2}$ Eq. \eqref{agents_group_trust_example2} with the help of agent-group RNE trust in Eq. \eqref{agent_group_trust} correspondingly. 

\begin{equation}
\begin{split}
    \mathop{\mathbb{T}(R_1||R_{G_1})} = & 0.4886 \cdot \log \frac{0.4886}{0.4869} + 0.4545 \cdot \log \frac{0.4545}{0.4657} + \\
    & 0.0568 \cdot \log \frac{0.0568}{0.0474} = 0.000915
\label{agents_group_trust_example1}
\end{split}
\end{equation}
\begin{equation}
\begin{split}
    \mathop{\mathbb{T}(R_1||R_{G_2})} = & 0.4886 \cdot \log \frac{0.4886}{0.5507} + 0.4545 \cdot \log \frac{0.4545}{0.3977} + \\
    & 0.0568 \cdot \log \frac{0.0568}{0.0516} = 0.007674
\label{agents_group_trust_example2}
\end{split}
\end{equation}

Here, since $\mathop{\mathbb{T}(R_1||R_{G_1})} < \mathop{\mathbb{T}(R_1||R_{G_2})}$ the trust of robot $R_1$ on the group $R_{G_1}$ is higher than its trust on the group $R_{G_2}$ because robot $R_1$ share similar needs with the average needs distribution of the group $R_{G_1}$.

\begin{proCoro}[Trust between Groups]
For certain state s$_1$ $\in$ S, supposing two groups R$_{G_1}$ and R$_{G_2}$ have the current needs matrix (N$_{11}$, \dots, N$_{1i}$) and (N$_{21}$, \dots, N$_{2i}$), Eq. \eqref{group_needs}. Considering each group member's needs vector ($n_{1k_1/2k_1}$, \dots, $n_{1k_j/2k_j}$) consists of different needs elements with corresponding weights W($w_1$, \dots, $w_j$). Let D$_{N1}$ (D$_{11}$, \dots, D$_{1j}$) and D$_{N2}$ (D$_{21}$, \dots, D$_{2j}$) present the agents needs' distribution of two groups. The RNE value (Trust) from group one to two in the current scenario can be defined as Eq. \eqref{trust}.
\begin{equation}
\label{group_needs}
\left[
\begin{matrix}
N_{11/21} \\ 
N_{12/22} \\ 
\vdots \\ 
N_{1i/2i} \\ 
\end{matrix} 
\right]
=
\left[
\begin{matrix}
n_{11_1/21_1} & n_{11_2/21_2} & \cdots & n_{11_j/21_j} \\ 
n_{12_1/22_1} & n_{12_2/22_2} & \cdots & n_{12_j/22_j} \\ 
\vdots & \vdots & \ddots & \vdots \\ 
n_{1i_1/2i_1} & n_{1i_2/2i_2} & \cdots & n_{1i_j/2i_j} \\ 
\end{matrix} 
\right]
\end{equation}

\begin{equation}
\label{group_needs_distribution}
\begin{split}
\mathop{D_{N1/N2}} & =
\left[
\begin{matrix}
D_{11/21} \\ 
D_{12/22} \\ 
\vdots \\ 
D_{1i/2i} \\ 
\end{matrix} 
\right]
=
\left[
\begin{matrix}
N_{11/21} \\ 
N_{12/22} \\ 
\vdots \\ 
N_{1i/2i} \\ 
\end{matrix} 
\right]_{i \times j}
. \times \\
&
\left[
\begin{matrix}
w_1 \\ 
w_2 \\ 
\vdots \\ 
w_j \\ 
\end{matrix} 
\right]^T
./
\left[
\begin{matrix}
\sum_{k=1}^{i} (N_{1i_k/2i_k} \cdot W_k)
\end{matrix}
\right]_{j \times 1}
\end{split}
\end{equation}

\begin{equation}
    \mathop{\mathbb{T}(D_{N_1}||D_{N_2})} = \sum_{k=1}^{j} D_{N_{1_k}} \cdot \log \frac{D_{N_{1_k}}}{D_{N_{2_k}}}
\label{trust}
\end{equation}
\end{proCoro}

\paragraph*{Example 3}
\label{example3}
Let us calculate the RNE trust value from group R$_{G_1}$ to R$_{G_2}$ based on Example 2 after ceiling the needs distributions to two decimals.
\begin{equation}
\begin{split}
    \mathop{\mathbb{T}(R_{G_1}||R_{G_2})} = & 0.48 \cdot \log \frac{0.48}{0.55} + 0.46 \cdot \log \frac{0.46}{0.39} + \\
    & 0.04 \cdot \log \frac{0.04}{0.05} = 0.0007, \\
    \mathop{\mathbb{T}(R_{G_2}||R_{G_1})} = & 0.55 \cdot \log \frac{0.55}{0.48} + 0.39 \cdot \log \frac{0.39}{0.46} + \\
    & 0.05 \cdot \log \frac{0.05}{0.04} = 0.0094.
\label{groups_trust_example}
\end{split}
\end{equation}

From this example, we can see that the robot group $R_{G_1}$ has a significantly higher level of trust on the group $R_{G_2}$ than the vice versa. This aligns with the intended social needs and motivations for intelligent agents (robots) where each robot views another robot from its own perspective rather than grounding on a common reference.

According to the above definitions and examples, we can see that RNE assesses the trust levels of the specific agent or team to others, which stands on the current individual or group perspective analyzing the needs' similarity from itself to others. It means that different agents or groups might have distinct RNE values or trust levels based on their current needs, knowledge of the environment and other agents, and future interests, much as humans do. In the real world, we usually assign the more challenging tasks to the agents or groups with reliable capabilities and stable performance, and vice versa for simpler tasks where grouping of agents does not need to account for trust within the group. However, most real-world tasks are challenging and has varying spatio-temporal difficulty levels. Therefore, grouping agents with similar corresponding capabilities and needs to the related mission is crucial in practical applications. 

\section{Heterogeneous Multi-Robot Systems}
\label{4}

We design a robot-aided urban search and rescue (USAR) mission to implement and illustrate the heterogeneous multi-robot grouping concept and corresponding algorithms. The mission is to retrieve (or rescue) as many resources (e.g., victims) as possible from the task area where the resources are present. In our scenarios, a set of robots will be available. Each robot is classified as one of the following: \textit{Carrier}, \textit{Supplier}, and \textit{Observer}. Each type of robot has a specific role and functionality \cite{yang2020needs}. Multiple robots cooperate to fulfill this rescue mission within a limited time \cite{luo2019multi}. When they form a group, there must be robots from each of these three categories, making the group heterogeneous in terms of functionalities.

\begin{figure}[tbp]
\centering
\includegraphics[width=0.49\textwidth]{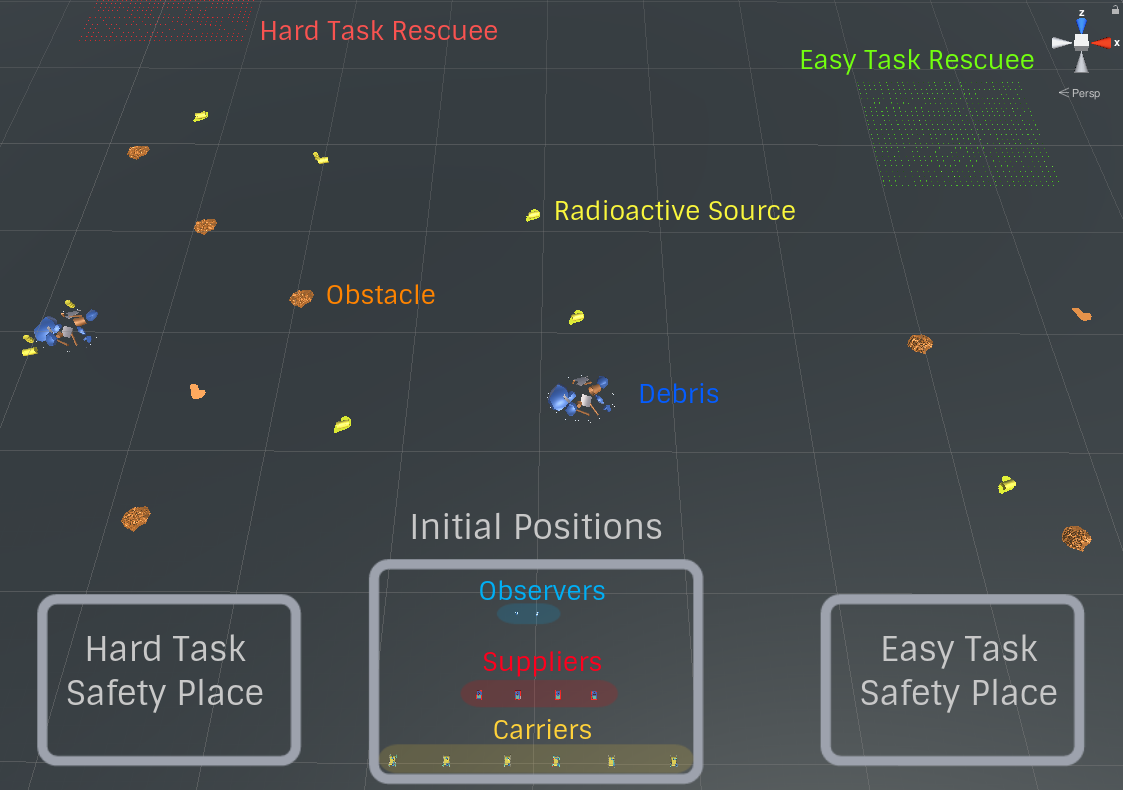}
\caption{Illustration of the simulation of a post-nuclear leak rescue mission.}
\label{fig: sim}
\end{figure}

This USAR mission is executed in multiple rounds by the heterogeneous MRS. Each round, the entire group will divide into two subgroups based on certain principles, executing the corresponding tasks.
There are two kinds of adversaries (debris and radioactive resources) randomly located in our post-nuclear leak rescue mission environment. According to the task's difficulty, the distance from the initial position to the task and the number of  adversaries distributed in the task area are different. 
See Fig. \ref{fig: sim} for an illustration of the heterogeneous MRS in the USAR mission under study.

Considering the various individual agent's needs in different situations, we organize their needs based on \textit{Robot Needs Hierarchy} as follows:
\begin{itemize}
    \item Safety Needs: HP (Health Points), Speed, and Sensing Range (detecting adversaries);
    \item Basic Needs: Energy (battery);
    \item Capability Needs: Capacity for carrying, Resource for rescuing, and Observing Range (detecting rescuee and rescued objects);
    \item Teaming Needs: The number of rescuees or rescued objects (Group Utility).
\end{itemize}

In our USAR mission, we assume that the number of \textit{Carrier}, \textit{Supplier} and \textit{Observer} are $x$, $y$ and $z$ ($x, y, z \in Z^+$). According to Section \ref{3} and \ref{4} discussion, we define their \textit{Needs Space}\footnote{The needs space has 7 variable, i.e., j=7 in Eqs.~\eqref{agents_trust}, \eqref{agent_group_trust}, and \eqref{trust}.} as Eqs. \eqref{carrier_needs}, \eqref{supplier_needs}, and \eqref{observer_needs}, respectively.
\begin{eqnarray}
    Carrier := N_c(hp_c, v_c, sen_c, eng_c, res_c, cap_c, obs_c);~~ \label{carrier_needs} \\
    Supplier := N_s(hp_s, v_s, sen_s, eng_s, res_s, cap_s, obs_s); \label{supplier_needs} \\
    Observer := N_o(hp_o, v_o, sen_o, eng_o, res_o, cap_o, obs_o). \label{observer_needs}
\end{eqnarray}
Here, 
\begin{itemize}
\item $hp$, $v$, and $sen$ represent the agent's safety needs like health points, velocity, and sensing range separately;
\item $eng$ represents the agent's basic energy needs such as battery level;
\item $cap$, $res$, and $obs$ represent the amount of the agent's carrying capacity and rescuing resources, and observing range for searching correspondingly.
\end{itemize}

When agents pass through the radioactive area, their health will be damaged continuously. For the safety needs, the individual needs should guarantee that the robot has enough HP supporting its normal function. From another perspective, through sensing perception and speed adaptation, they should avoid the collision between agents and adversaries like obstacles (debris) and radioactive sources. Similarly, they also need to satisfy corresponding energy fulfilling tasks \cite{parasuraman2012energy}. Otherwise, it will return to the resting place for recharging its energy or refreshing its HP.

The individual robot should represent corresponding abilities for the task to satisfy the mission's capability needs, especially for the heterogeneous agents' cooperation. Our USAR mission assumes that saving one rescued object will cost one space (capacity) and one rescuing resource. Besides, each group member's speed will decrease by 30\% without an observer involving the group, and group members can share their resources in the specific task.


\section{RNE Trust-based Multi-Robot Cooperation}
\label{5}

Supposing we have two tasks in our USAR mission, which means that the entire group needs to divide into two subgroups fulfilling the corresponding task. These two tasks have different difficulty levels - easy and hard. Compared to the easy task, the hard task has more obstacles (debris) and radioactive sources in the disaster area, which means that agents might cost more energy avoiding obstacles and more HP resisting radiation emitting from radioactive resources. Besides, agents will spend more time and energy executing the challenging task because of the longer distance.

For the multi-robot grouping, we organize the group members and implement the \textit{bottom-up} mechanism instead of partition without classification and hierarchy. More specifically, in the three homogeneous groups -- \textit{Carrier}, \textit{Supplier} and \textit{Observer}, we first consider separating them into two subgroups based on different combinations $C_x^{\frac{x}{2}}$, $C_y^{\frac{y}{2}}$, and $C_z^{\frac{z}{2}}$, respectively. Then, we can always find a combination with the most significant RNE value difference between those subgroups in the corresponding homogeneous group. Supposing the three groups' final combinations are $G_c(Car_1, Car_2)$, $G_s(Sup_1, Sup_2)$, and $G_o(Obs_1, Obs_2)$, using the same approach, we can merge the subgroups of carrier and supplier into a new combination $G_{cs}(CS_1, CS_2)$ recursively. Until we get the final group partition $G_{cso}(CSO_1, CSO_2)$, we finish the entire RNE trust-based grouping. The whole process can be represented as the main RNE trust-based multi-robot grouping algorithm in Alg.~\ref{alg:rne_group}, with supporting functions and algorithms in Alg.~\ref{alg:asc} (obtaining hierarchical subgroup combinations), \ref{alg:grc} (calculating group's RNE trust values), Alg. \ref{alg:ndc} (calculating needs distribution vector of a robot group), and Alg. \ref{alg:ds} (iteratively solving deadlocks in robot's task assignments).

\begin{algorithm}[t]
\SetAlgoLined
\SetKwData{Left}{left}\SetKwData{This}{this}\SetKwData{Up}{up}
\SetKwFunction{Union}{Union}\SetKwFunction{FindCompress}{FindCompress}
\SetKwInOut{Input}{Input}\SetKwInOut{Output}{Output}
\SetNlSty{textrm}{}{:}
\SetKwComment{tcc}{/*}{*/} 

\scriptsize{

\Input{
		Carrier, Supplier, and Observer group state matrixes: $A_o$, $A_s$, and $A_o$. The number of tasks: $2$.
       }
\Output{
		The maximum RNE difference group partition ($cso_1$, $cso_2$).
}
\BlankLine
initialization\;
memory data set $V_{c/s/o/cs/cso}$ = \{$\varnothing$\}; \\
$m$ = $A_{c/s/o}$.Count / 2; \\
    \For{each i $\in$ $ASC(A_{c/s/o}, m)$}
    {$V_{c/s/o}$.Add($GRC(i[0],i[1])$)}
$G_{c/s/o}$ = Max($V_{c/s/o}$); \\
    \For{each i $\in$ $G_c$}
    {\For{each j $\in$ $G_s$}
    {$V_{cs}$.Add($GRC(i,j)$)}}
$G_{cs}$ = Max($V_{cs}$); \\
    \For{each i $\in$ $G_{cs}$}
    {\For{each j $\in$ $G_o$}
    {$V_{cso}$.Add($GRC(i,j)$)}}
\Return $G_{cso}$ = Max($V_{cso}$).
} 
\caption{\footnotesize{RNE Trust-based Multi-robot Grouping}}
\label{alg:rne_group}
\end{algorithm}

\begin{algorithm}[t]
\SetAlgoLined
\SetKwData{Left}{left}\SetKwData{This}{this}\SetKwData{Up}{up}
\SetKwFunction{Union}{Union}\SetKwFunction{FindCompress}{FindCompress}
\SetKwInOut{Input}{Input}\SetKwInOut{Output}{Output}
\SetNlSty{textrm}{}{:}
\SetKwComment{tcc}{/*}{*/} 

\scriptsize{

\Input{
		Group members' state list $A$;
		The partition point: $m$.
       }
\Output{
        All subgroup combination list $S_A$.
}
\BlankLine
initialization\;
    memory data set $agent$, $V$, $D$ = \{$\varnothing$\} \\
    \If{$A$.Count $==$ $m$ And $m \neq 1$}
    {\Return $S_A$.Add($A$)}
    \ElseIf{$m == 1$}    
    {
        \For{each $i$ $\in$ $A$}
        {
            $V$.Add($i$); \\
            $S_A$.Add($V$); \\
            $V=Null$.
        }
        \Return $S_A$;
    }
\For{each $i$ $\in$ ($A$.Count - $m$)}
{
    $agent$ = $A[0]$; \\
    $A$.Remove($0$); \\
    $D$ = $ASC$($A$, $m-1$); \\
    \For{each $j$ $\in$ $D$.Count}
    {
        $V$.Add($agent$); \\
        $V$.Add($D[j]$); \\
        $S_A$.Add($V$); \\
        $V=Null$.
    }
}

\Return $S_A$.
} 

\caption{\footnotesize{All Subgroup Combinations (ASC)}}
\label{alg:asc}
\end{algorithm}

\begin{algorithm}[t]
\SetAlgoLined
\SetKwData{Left}{left}\SetKwData{This}{this}\SetKwData{Up}{up}
\SetKwFunction{Union}{Union}\SetKwFunction{FindCompress}{FindCompress}
\SetKwInOut{Input}{Input}\SetKwInOut{Output}{Output}
\SetNlSty{textrm}{}{:}
\SetKwComment{tcc}{/*}{*/} 

\scriptsize{

\Input{
		Group 1 and 2 state matrixes $A_1$ and $A_2$.
       }
\Output{
        The RNE value from group 1 to 2 $RNE_{12}$.
}
\BlankLine
initialization\;
    memory data set $D_{1/2}$ = \{$\varnothing$\} \\
    $D_1$ = $NDC$($A_1$.needs, $A_1$.weights); \\
    $D_2$ = $NDC$($A_2$.needs, $A_2$.weights); \\
    \For{each i $\in$ $Max$($D_1$.Count, $D_2$.Count)}
    {
        $RNE_{12}~~+= D_1[i]$ $\times$ $\log(D_1[i] / D_2[i])$
    }
\Return $RNE_{12}$
} 

\caption{\footnotesize{Groups RNE Calculation (GRC)}}
\label{alg:grc}
\end{algorithm}

\begin{algorithm}[t]
\SetAlgoLined
\SetKwData{Left}{left}\SetKwData{This}{this}\SetKwData{Up}{up}
\SetKwFunction{Union}{Union}\SetKwFunction{FindCompress}{FindCompress}
\SetKwInOut{Input}{Input}\SetKwInOut{Output}{Output}
\SetNlSty{textrm}{}{:}
\SetKwComment{tcc}{/*}{*/} 

\scriptsize{

\Input{
		Group members' needs and weight matrix: $N$ and $W$.
       }
\Output{
		The group needs' distribution vector $D$.
}
\BlankLine
initialization\;
memory data set $V$, $D$ = \{$\varnothing$\} \\
    \For{each i $\in$ $N$.Count}
    {
        $V$.Add($N[i]$ $\times$ $W[i]$)
    }
    \For{each i $\in$ $V$.Count}
    {
        $D$.Add($V[i]$ / $V$.Sum())
    }   
\Return $D$.Sort()
} 

\caption{\footnotesize{Needs' Distribution Calculation (NDC)}}
\label{alg:ndc}
\end{algorithm}

\begin{algorithm}[t]
\SetAlgoLined
\SetKwData{Left}{left}\SetKwData{This}{this}\SetKwData{Up}{up}
\SetKwFunction{Union}{Union}\SetKwFunction{FindCompress}{FindCompress}
\SetKwInOut{Input}{Input}\SetKwInOut{Output}{Output}
\SetNlSty{textrm}{}{:}
\SetKwComment{tcc}{/*}{*/} 

\scriptsize{

\Input{
		Sorted collision agents' state list $C$ based on agent's needs.
       }
\Output{
		Solving the deadlock.
}
\BlankLine
\% Detect the deadlock. \\
\While{length($C$) $\neq$ $0$ And $\Delta C$.velocity $ == 0$ }
{   
    $i$++; \\
    \% Switch the execution order from the high to the low priority agent. \\
    $C_n$ = $Swap$($C$, $0$, $C[i]$); \\
    \% Execute the new collision list in the collision avoiding planning. \\
    CollisionAvoiding($C_n$); \\
    \% Check whether the deadlock is existing. \\
    $DSM$($C_n$).
}
$i = 0$; \\
$C_n =$ \{$\varnothing$\}; \\
\Return
} 

\caption{\footnotesize{Deadlock Solving Mechanism (DSM)}}
\label{alg:ds}
\end{algorithm}

The multi-robot grouping implementation is based on our previous work in {\it Self-Adaptive Swarm System} (SASS) \cite{yang2020hierarchical}. Here, the individual agent selecting a suitable group relies on the trust (RNE) calculation, which means that agents with similar needs levels will form a group executing the corresponding task. Besides, through the {\it Negotiation-Agreement Mechanism} and route planning, agents can efficiently avoid conflicts and collisions between agent to agent and agent to static obstacles (debris). More specifically, for the SASS \textit{Collision Avoidance} planning, the agent with the highest priority in the collision list will move first, and the rest of the agents stop until it out of the list. Then, they keep on executing the process recursively until the collision list is empty. Especially for this deadlock solving problem, when all the collision list agents can not move, we build a dynamic priority-based switching mechanism providing each agent has a fair opportunity to move until solving the deadlocks (see Alg. \ref{alg:ds}).

\section{Evaluation through Simulation Studies}

Considering the cross-platform, scalability, and efficiency of the simulations, we chose the ``Unity'' game engine to simulate the USAR mission. In our simulations, we consider six carriers, four suppliers, and two observers in total which separated into two equally-numbered subgroups executing two complex tasks in a post-nuclear rescue mission to rescue the resources and valuable items as much as possible within a limited time period. We compare the RNE trust-based grouping method with other methods from the literature.

\subsection{Performance Metrics}

We consider the following performance metrics in this study for a comparative analysis.

\paragraph*{No. of Rescuees} We calculate the total number of resources the robot groups have rescued in both the tasks  in the USAR mission for ten minutes per trial.

\paragraph*{Energy Cost Per Rescuee} This is the total battery level spent by all robots divided by the total number of rescuees. From the agent basic needs perspective, we calculate the energy cost per rescuee for each category task and analyze the integrated energy cost per rescuee per trial.

\paragraph*{HP Cost Per Rescuee} This is the total Health points used by all robots divided by the total number of rescuees. Here, we analyze the HP cost per rescuee performance similar to the {\it Energy Cost Per Rescuee}.



\subsection{Simulation Studies}

Until the timeout is reached, multiple rounds of rescue mission is executed. After each round, when all agents are back to the initial position, the whole group members will regroup based on agents' current grouping principle, then distributed to the corresponding task continuing the mission. From the individual perspective, if it passes through the radioactive area, its HP will cost 0.0003\% per step, and we assume that each agent costs 0.0045\% energy (battery point) for every moving step. Besides, to satisfy the agent's safety and basic needs, we assume that if the individual battery (energy) or HP level is below 30\%, it needs to go to the pre-defined rest position for recharging energy (or for refreshing its health HP) 30 seconds, then back to the initial position waiting for the next round to start.  

In the experiments, the agent's HP level affects its speed, sensing, and observing range, which means that the lower its HP, the smaller its velocity, sensing, and observing range. Besides, without observer involvement, each group member's speed will decrease by 30\%. Moreover, each group member can share their resources and information between a heterogeneous subgroup to achieve the specific task cooperatively in every round. We present the needs' relationships between \textit{Carrier}, \textit{Supplier}, and \textit{Observer} as follows:

\begin{itemize}
\item {\it Safety Needs}: $hp_s = hp_c = 2hp_o$; $v_c = v_s = 1.5v_o$; $sen_o = 6sen_c = 6sen_s$.
\item {\it Basic Needs}: $eng_s = eng_c = 1.5eng_o$;
\item {\it Capability Needs}: $cap_c = 6cap_s$, $cap_o = 0$; $res_s = 10res_c$, $res_o = 0$; $obs_o = 1000obs_c = 1000obs_s$.
\end{itemize}

Moreover, each subgroup (i.e., each task) has one observer, two suppliers, and three carriers. We conduct ten simulation trials for each principal case (method) in the whole experiment. In every case, we use the same ten different initial battery (energy) and HP levels sampled from two Gaussian distributions with means of $80\%$ and $90\%$, and standard deviations of $20\%$ and $10\%$ respectively.

\subsection{Compared methods from the literature}

To evaluate our RNE Trust model for grouping, we implement the state-of-the-art energy-based (ENG) method \cite{yang2020hierarchical} and use the distance-based (DIS) grouping approach as the baseline in the performance comparison. Besides, we combine the agent's HP level and distance (HP$_{DIS}$) as another grouping principle applied in the experiments. We discuss the specific details for each case (method) as follows:

\paragraph*{$DIS$ - Distance-based} Each agent will compare the distances between its initial position and the goal points of two tasks, then selects the nearest one forming a group to execute. They continue to form a group until the necessary count of robots has reached for that group. This is a typical mode of multi-robot grouping used in the literature. For instance, in \cite{parasuraman2018multipoint}, the authors used shortest distance to the task as the measure to determine the group membership. 

\paragraph*{$ENG$ - Energy-based} According to our previous work in \cite{yang2020hierarchical,yang2019self}, where the agents with the same category will sort based only on their current energy (battery) levels at the initial positions, then divide into two parts with equal numbers. By selecting the high or low parts in each category to form the corresponding two heterogeneous subgroups, we assign the higher energy group to the challenging task and the lower one to the easy one, respectively.

\paragraph*{$HP_{DIS}$ - Health points combined with the distance based} Similar to the energy-based grouping, here we use the health level of the robots. First, consider sorting the robots in each category (type) based on their current health points (HP) at the initial position. For the agents with the same HP level, we prioritize them by the $DIS$ approach.

\paragraph*{RNE Trust} By calculating the proposed RNE values between agents at the initial position, the whole group classifies as two subgroups with maximum trust level difference between each other. It means that the subgroup gathers agents with similar higher capabilities and needs levels. Then, the subgroup with higher trust level within the group executes the hard task, and the lower trust group fulfills the easy task.

\begin{figure}[t!]
\begin{center}
    \subfigure[Group Utility (Teaming Needs) Analysis]{
    \begin{minipage}[t]{1\linewidth}
    \centering
 \includegraphics[width=0.94\textwidth]{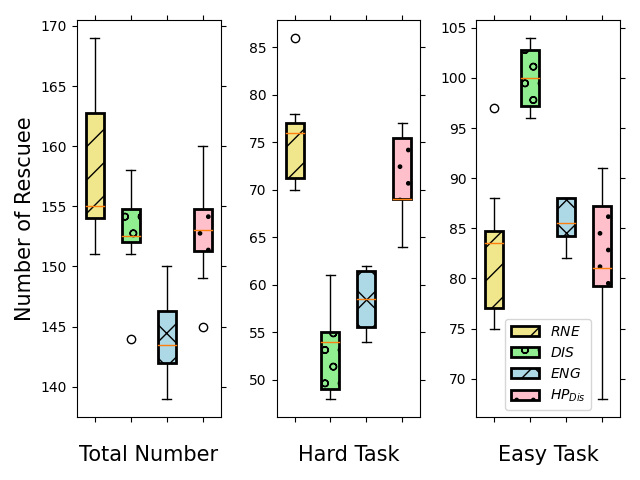}
 \label{fig: eahc}
 \end{minipage}}
  \vspace{-2mm}
    \subfigure[Energy Cost Analysis]{
    \begin{minipage}[t]{1\linewidth}
    \centering
    \includegraphics[width=0.94\textwidth]{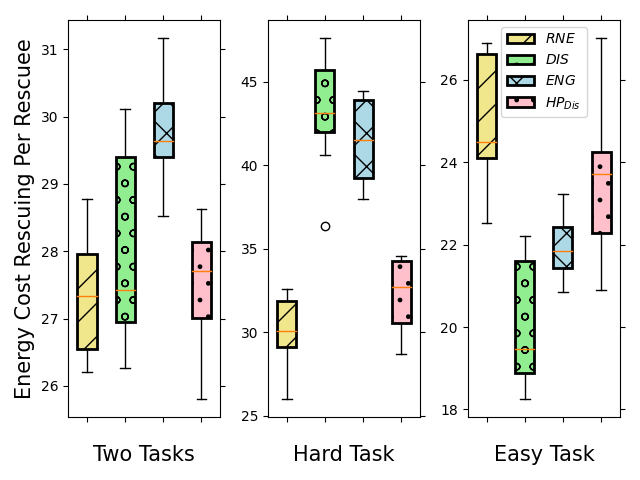}
    \label{fig: kpmel}
    \end{minipage}}
    \vspace{-2mm}
 \subfigure[Health (HP) Cost Analysis]{
    \begin{minipage}[t]{1\linewidth}
    \centering
    \includegraphics[width=0.94\textwidth]{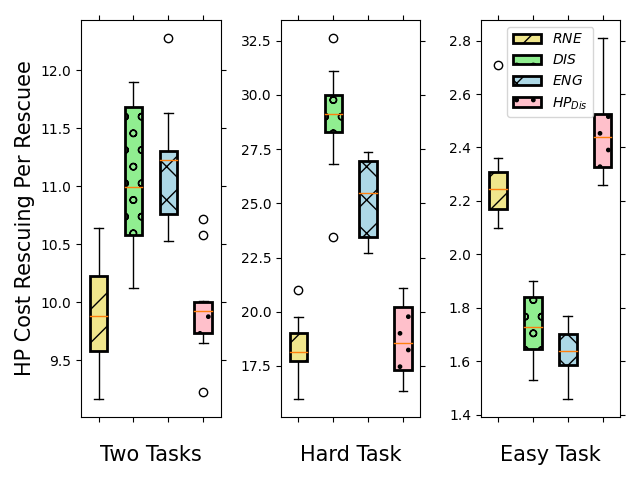}
    \label{fig: kpmhc}
    \end{minipage}}
\caption{\small{Performance comparison of different multi-robot cooperation (grouping) models in USAR missions with two difficulty tasks.}}
\label{fig:interactive_experiment}
\vspace{-4mm}
\end{center}
\end{figure}

\section{Results and Discussion}

Figure~\ref{fig:interactive_experiment} show the final results of all the methods compared in this study showing all three performance metrics. The experiment demonstration video is available at our website \url{http://hero.uga.edu/research/trust/}.
Through Fig. \ref{fig: eahc}, we can notice that RNE has the best performance for the total number of rescuees than other methods. More specifically, RNE presents more advantages in the challenging task but inferior results in the easy task, where pure distance-based group formation is highly effective. The RNE trust-based grouping mechanism balances various needs between agents and organizes more reliable and stable groups for achieving the tough or emergent task, and uses the lower trusted agents for easy task.

From the perspective of optimizing the efficiency of system resources, Figs. \ref{fig: kpmel} and \ref{fig: kpmhc} show that RNE essentially decreases the energy and HP cost rescuing per rescuee, especially in the challenging task. The RNE trust model helps the system reassign the resources and gathers agents with similar capabilities, needs, and interests to fulfill the suitable tasks, which improves the system performance and lets each group member's abilities be fully utilized. Like human society, the multi-robot system can make the best possible use of resources and materials based on trust-based cooperation models.

Comparing with the natural intelligent agent, when the artificial intelligence (AI) agent becomes more advanced and smart, it also represents more complex, multilayered, and diverse needs in evolution such as individual security, health, friendship, love, respect, recognition, and so forth. When we consider intelligent agents, like robots, working as a team or cooperating with human beings, organizing their needs building certain reliable and stable relationships such as trust is a precondition for robot-robot and human-robot collaboration in complex and uncertain environments \cite{yang2020needs}.

\section{Conclusion}

In this paper, we introduce a general agent trust model based on \textit{Relative Needs Entropy (RNE)} to measure and analyze the trust levels between agents and groups, representing the similarity of their diverse needs in a specific situation for heterogeneous multi-robot cooperation. Then, we illustrate how the RNE trust can be used in multiagent decision-making applications. Specifically, we propose an RNE trust-based effective heterogeneous multi-robot cooperation method to form multiple robot groups based on trust levels within the groups. The proposed model is evaluated through extensive simulations under different difficulty tasks in a post-nuclear radiation leak-like urban search and rescue scenario. We also developed a dynamic priority switching mechanism to solve conflicts in multi-robot cooperation. 

The experimental analysis showed that the RNE trust-based grouping model outperformed state-of-the-art energy-based and distance-based methods in maximizing group utilities and represented lower system costs.
Trust based on relative needs distributions presents opportunities for improvements and interesting applications. Ultimately, we envision a harmonious team of robots in future multi-robot missions in which each robot values trust on each other robot.




\bibliographystyle{IEEEtran}
\bibliography{references}



\end{document}